\documentclass[twoside,english,nofootinbib,preprintnumbers,aps,onecolumn,notitlepage,11pt]{revtex4}
\usepackage{euscript,amssymb}
\usepackage{amsthm}
\usepackage{graphicx}
\usepackage{amsfonts}
\usepackage{amsmath}
\usepackage{amssymb}
\usepackage{fancyhdr}
\usepackage{epsfig}
\usepackage{color}
\usepackage{soul}
\usepackage{enumitem}
\usepackage{calc}
\usepackage{enumerate}
\usepackage{afterpage}
\usepackage{multirow}
\usepackage[papersize={8.5in,11in}]{geometry}

\allowdisplaybreaks

\newcommand{\be}{\begin{equation}}
\newcommand{\ee}{\end{equation}}

\begin{document}
%%%%%%%%%%%%%%%%%%%%%%%%%%%%%%%%%%%%%%%%%%%%%%%%%%%%%%%%%%%%%%%%%%%%%%%%%%%%%%%%%%%%%
\title{A generalization of the quadruple formula for the energy of gravitational radiation in de Sitter spacetime}
%%%%%%%%%%%%%%%%%%%%%%%%%%%%%%%%%%%%%%%%%%%%%%%%%%%%%%%%%%%%%%%%%%%%%%%%%%%%%%%%%%%%%
\author{Denis Dobkowski-Ry{\l}ko}
	\email{Denis.Dobkowski-Rylko@fuw.edu.pl}
	\affiliation{Faculty of Physics, University of Warsaw, ul. Pasteura 5, 02-093 Warsaw, Poland}
	%\author{Wojciech Kami\'nski}
	%\email{Wojciech.Kaminski@fuw.edu.pl}
	%\affiliation{Faculty of Physics, University of Warsaw, ul. Pasteura 5, 02-093 Warsaw, Poland}

\author{Jerzy Lewandowski}
	\email{Jerzy.Lewandowski@fuw.edu.pl}
	\affiliation{Faculty of Physics, University of Warsaw, ul. Pasteura 5, 02-093 Warsaw, Poland}
\begin{abstract}
We study gravitational radiation produced  by time changing matter source in de Sitter spacetime. We consider a cosmological Killing horizon instead of the conformal boundary used in the radiation theory in the Minkowski spacetime.  The energy of the radiation passing through the horizon is derived. Our result  takes the form of a generalized quadruple formula expressed in terms of the mass and pressure quadruple moments and is written explicitly up to the first order in $\sqrt{\Lambda}$. The zeroth order term recovers the famous Einstein's quadruple formula obtained for the perturbed Minkowski spacetime, whereas the first order term is a new correction.
%{\color{black} We also calculate the first correction to the energy coming from the presence of the positive cosmological constant.}}
\end{abstract}

\date{\today}

\pacs{???}

\maketitle

\section{Introduction}
%The goal of this work is to calculate the energy carried by gravitational waves in the $\Lambda>0$ case and the natural way to approach it is to study the perturbed de Sitter metric.
We study the linearized Einstein's equations to investigate the radiation coming from the isolated systems. Our goal is to calculate the energy carried by gravitational waves in the $\Lambda>0$ case and the natural way to approach it is to study the perturbed de Sitter metric. For the vanishing cosmological constant $\Lambda$ one usually considers small perturbations in Minkowski spacetime, where the suitable framework to analyze gravitational radiation is that of the conformal boundary often called scri plus and denoted by $\mathcal{I}^+$. A generalization of the notion of the Minkowski spacetime $\mathcal{I}^+$  to the de Sitter spacetime is ambiguous and depends on properties we want to preserve. If this is going to be the conformal boundary, then the obtained $\mathcal{I}^+$ is no longer a null surface. Instead of being null, it is spacelike. On the other hand, if we insist that a generalized $\mathcal{I}^+$ in de Sitter spacetime is a null surface, then a good candidate is a cosmological horizon \cite{ABK1, ABK2}. We implement that latter approach to the radiation in de Sitter spacetime. Our strategy is as follows. We consider the gravitational radiation produced by a compact time-changing source in the slow motion approximation and assume that the cosmological constant $\Lambda$ is very small. Due to the latter assumption the cosmological horizon becomes very distant from the source.  Next we apply the Wald-Zoupas \cite{WZ} and Chandrasekaran-Flanagan-Prabhu \cite{CFP} theory of radiation through the null surfaces. To make it applicable, we need to introduce a gauge in which the horizon is still a null surface with respect to the perturbed geometry. Alternatively, one can interpret such procedure as deforming the cosmological horizon such that given the original perturbation of the spacetime it remains null.  Finally, we calculate the energy, that corresponds to the horizon forming Killing vector, to the second order in the perturbation and  to the first order in  $\sqrt{\Lambda}$. We compare the zeroth order term in  $\sqrt{\Lambda}$ with the standard Einstein's quadrupole formula (find that they both agree), and calculate the first order term in $\sqrt{\Lambda}$ as the correction coming from the cosmological constant.

There are several related papers or works in the literature, that inspired the current one.  Most importantly,  a generalization of the quadrupol formula for de Sitter spacetime as the background spacetime was derived by using the spacelike conformal boundary of de Sitter spacetime in \cite{ABK3}. Here we calculate the radiation flux through a different surface, hence the results may be different, except for  the total integrals along the surfaces that agree according to the symplectic theory on the space of solutions to Einstein's equations.  A  description  of the radiation through a general non-expanding horizon similar to the current one  was studied in \cite{AKKL1, AKKL2}, however a significant role played the assumption that the expansion of the perturbed horizon vanishes asymptotically in the future. That assumption is motivated by an example of a black hole that is a final product of a binary coalescence. In such case, it follows that the expansion has to vanish everywhere on the horizon in the linear order.  The vanishing of the expansion of a perturbed NEH in a generic case may be achieved by a suitable gauge fixing by similar arguments that of \cite{WH} which considered Killing horizons. However, in our case the argument from \cite{WH} may not be applied, because the suitable MOTS operator in the case of the Killing horizon in de Sitter spacetime is not invertible. This is why the considered case here differs from the one in \cite{WH}. Similarly, results found in \cite{KL1} are not applicable to the radiation coming from a compact source.

In the first part of our work we will recall some of the results obtained by Ashtekar et al. in \cite{ABK3}. We will use the solution to linearized Einstein's equations as well as the notion of mass and pressure quadruple moments introduced in \cite{ABK3} and find the generalized quadruple formula passing through the Killing horizon for the energy carried by gravitational radiation.

\begin{figure}[h]
    \centering
    \includegraphics[width=0.7\textwidth]{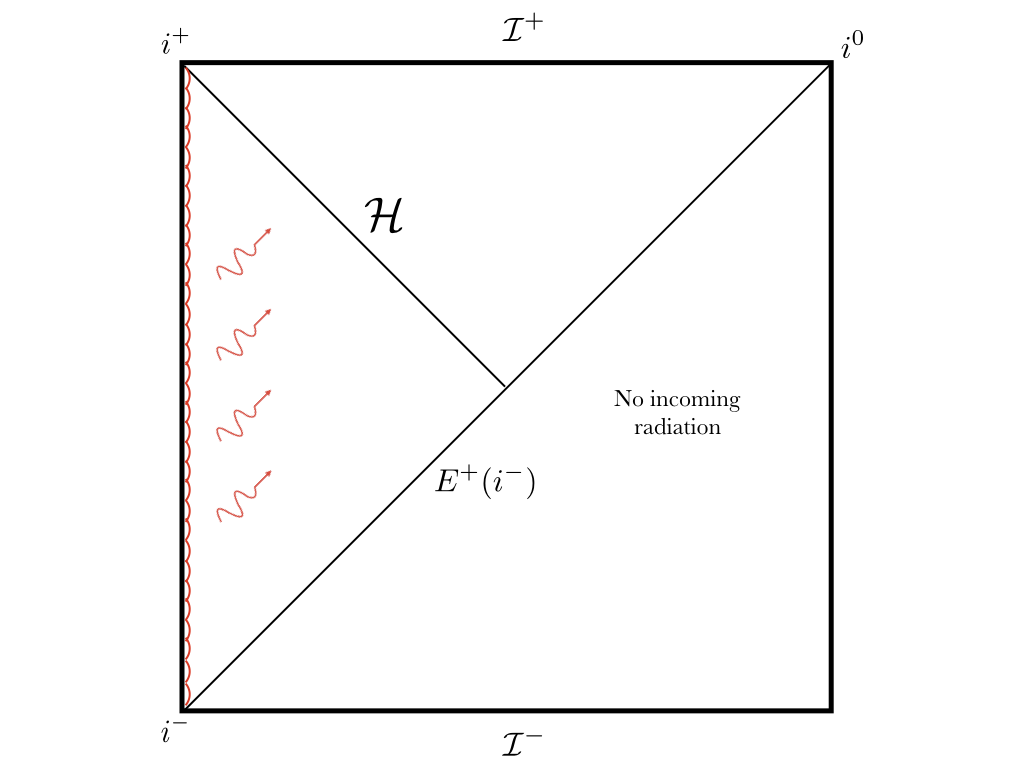}
    \caption{A time changing matter source emitting gravitational waves. The spatial size of the system is uniformly bounded in time, whereas its causal future covers the future Poincar\'e patch (triangle with vertices at $i^-$, $i^0$ and $i^+$). We assume no incoming radiation across the past boundary  $E^+(i^-)$ (future  event horizon of $i^-$) of the future Poincar\'e patch. $\mathcal{H}$ denotes the cosmological horizon.}
    \label{pdiag}
\end{figure}

\vspace{5mm}
Consider a time changing matter source in de Sitter spacetime emitting gravitational waves whose spatial size is uniformly bounded in time (see Fig. \ref{pdiag}). Its causal future covers the so called future Poincar\'e patch, which we coordinatize by $(\eta,x,y,z)$. The background de Sitter metric $\bar g_{\alpha\beta}$ takes the following form:
\begin{align}\label{etametric}
\bar g_{\alpha\beta} dx^{\alpha}dx^\beta =\tfrac{1}{H^2\eta^2}\mathring g_{\alpha\beta} dx^{\alpha}dx^\beta= \tfrac{1}{H^2\eta^2}(-d\eta^2+dx^2+dy^2+dz^2),
\end{align}
where $H:= \sqrt{\Lambda/3}$ is the Hubble parameter and $\Lambda$ is the cosmological constant. On the future Poincare patch the comoving spatial coordinates $(x,y,z)$ span $\mathbb{R}$ whereas the conformal time coordinate $\eta$ takes values in the interval $(-\infty, 0)$.

To find the flux of energy radiated by an isolated system in the presence of the positive cosmological constant we considered the first order perturbations of de Sitter spacetime and use the solution to the linear Einstein's equations derived by Ashtekar et al. in \cite{ABK3}. Following their work we introduce the perturbed metric tensor $g_{\alpha\beta}$:
\begin{align}
g_{\alpha\beta}= \bar g_{\alpha\beta}+\epsilon \gamma_{\alpha\beta},
\end{align}
where $\epsilon$ is a smallness parameter. Next, it is convenient to consider trace-reversed metric perturbation:
\begin{align}
\bar \gamma_{\alpha\beta}:=\gamma_{\alpha\beta}-\tfrac{1}{2}\bar g_{\alpha\beta} \gamma,
\end{align}
and write the linearized Einstein's equations in the presence of a first order linearized source $T_{\alpha\beta}$ in the following form:
\begin{align}\label{LEE}
 \bar \Box \bar \gamma_{\alpha\beta}-2\bar\nabla_{(\alpha}\bar\nabla^\mu\bar\gamma_{\beta)\mu}+\bar g_{\alpha\beta}\bar\nabla^\mu\bar\nabla^\nu \bar \gamma_{\mu\nu} - \tfrac{2}{3}\Lambda(\bar\gamma_{\alpha\beta}-\bar g_{\alpha\beta}\bar\gamma) = -16\pi T_{\alpha\beta},
\end{align} 
where $\bar \nabla$ and $\bar \Box$ are the derivative and the d'Alembertian operators with respect to the de Sitter metric $\bar g_{\alpha\beta}$. Solution to (\ref{LEE}) have been derived and studied thoroughly in particular in \cite{Vega, ABK3} and in the next section we will just recall the main steps of its derivation.

Throughout this paper we use the following (abstract) index notation:
\begin{itemize}
\item Indices of the spacetime tensors are denoted by lower Greek letters: $\alpha,\beta,\gamma,... = 1,2,3,4$. 
\item Tensors defined on the cosmological slices $\eta=\text{const}$ carry indices denoted by lower Latin letters: $a,b,c, ... = 1,2,3$.
\item Tensors defined on the cosmological horizon $\mathcal{H}$ are denoted by lower Latin letters: $i,j,k, ... = 1,2,3$.
\item Capital Latin letters $A,B,C,...=1,2$ are used as the indices of tensors defined on the 2-dimensional space $S$ of null curves in $\mathcal{H}$.
\end{itemize}

\section{Retarded solution to linearized Einstein's equations\newline and its relation to quadruple moments}
In this section we will continue to follow Ashtekar et al. \cite{ABK3}, and introduce the trace-reversed, rescaled metric perturbation to solve the linearized Einstein's equations as well as define the mass and pressure quadruple moments, which will be used for the energy flux formula in the following part of our work. Unlike in \cite{ABK3}, not only the spatial part of the retarded solution but also the $\eta\alpha$-component will be written in terms of the quadruple moments and to accomplish that we use the gauge condition.

\subsection{Solving the linearized Einstein's equations}
Consider a vector field $\eta^\alpha$ normal to the $\eta=\text{const}$ hypersurface such that:
\begin{align}
\eta^\alpha \nabla_\alpha \eta =1.
\end{align}
Therefore, written in the introduced coordinates it takes a form:
\begin{align}
\eta^\alpha \partial_\alpha = \partial_\eta.
\end{align}
Denote by $n^\alpha$ the future pointing unit normal to those slices, namely:
\begin{align}
n^\alpha:= -H\eta\eta^\alpha.
\end{align}
Following \cite{ABK3} we use the following gauge condition:
\begin{align}\label{gauge1}
\bar\nabla^\alpha \bar\gamma_{\alpha\beta} = 2Hn^\alpha \bar \gamma_{\alpha\beta},
\end{align}
which significantly simplifies the linearized Einstein's equations (\ref{LEE}). Next define:
\begin{align}
\bar\chi_{\alpha\beta} := H^2\eta^2\bar \gamma_{\alpha\beta},
\end{align}
and introduce the following decomposition:
\begin{align}
\tilde \chi := (\eta^\alpha\eta^\beta+\mathring q^{\alpha\beta})\bar \chi_{\alpha\beta}, && \chi_\alpha :=\eta^\sigma\mathring {q_\alpha}^\beta \bar\chi_{\beta\sigma}, && \chi_{\alpha\beta}:=\mathring {q_\alpha}^\mu \mathring{q_\beta}^\nu \bar \chi_{\mu\nu},\\
 \mathcal{\tilde T}:= (\eta^\alpha\eta^\beta+\mathring q^{\alpha\beta})T_{\alpha\beta}, && \mathcal{T}_\alpha :=\eta^\sigma\mathring {q_\alpha}^\beta  T_{\beta\sigma}, && \mathcal{T}_{\alpha\beta}:=\mathring {q_\alpha}^\mu \mathring{q_\beta}^\nu T_{\mu\nu}.
\end{align}
where $\mathring q^{\alpha\beta}$ is the induced by the flat metric $\mathring g_{\alpha\beta}$, contravariant spatial metric on a cosmological slice $\eta=\text{const}$. The gauge conditions (\ref{gauge1}) rewritten using this decomposition take a form:
\begin{align}\label{gauge2}
\mathring D^a \chi_{ab} = \partial_\eta \chi_b - \tfrac{2}{\eta} \chi_b, && \mathring D^a\chi_a = \partial_\eta(\tilde \chi-\chi) - \tfrac{1}{\eta}\tilde \chi,
\end{align}
where $\mathring D$ is the derivative operator of the metric $\mathring q_{ab}$. As a consequence the linearized Einstein's equations ($\ref{LEE}$) break into three equations:
\begin{align}
\mathring \Box \Big(\tfrac{1}{\eta} \tilde \chi \Big) &= -\tfrac{16\pi }{\eta}  \mathcal{\tilde T},\\
\mathring \Box \Big(\tfrac{1}{\eta}  \chi_a \Big) &= -\tfrac{16\pi }{\eta} \mathcal{T}_a,\\
 \Big(\mathring \Box+\tfrac{2}{\eta} \partial_\eta \Big)\chi_{ab} &= -{16\pi }\mathcal{T}_{ab} \label{LEE3}.
\end{align}
We are particularly interested in totally spatial projection $\chi_{ab}$ which will be used for expressing the calculated energy flux in terms of the quadruple moments. Solving the above equations and assuming no incoming radiation across the past boundary of the future Poincar\'e patch yields:
\begin{align}
\tilde \chi(\eta,\vec x) &= 4\eta \int \frac{d^3\vec x'}{|\vec x-\vec x'|}\frac{1}{\eta_{{}_\text{Ret}}} \mathcal{\tilde T}(\eta_\text{ret},\vec x') \label{exactsolution1}\\
\chi_{a}(\eta,\vec x) &= 4\eta \int \frac{d^3\vec x'}{|\vec x-\vec x'|}\frac{1}{\eta_{{}_\text{Ret}}}\mathcal{T}_a(\eta_\text{ret},\vec x') \label{exactsolution2}\\
\chi_{ab}(\eta,\vec x) &= 4\int \frac{d^3 \vec x'}{|\vec x - \vec x'|} \mathcal{T}_{ab}(\eta_{_\text{Ret}},\vec x') + 4\int d^3\vec x' \int_{-\infty}^{\eta_{_\text{Ret}}} d\eta' \frac{1}{\eta'} \partial_{\eta'}\mathcal{T}_{ab}(\eta',\vec x')\label{exactsolution}
\end{align}
where $\eta_{_\text{Ret}}:=\eta-{|\vec x - \vec x'|}$. We will consider this solution on the cosmological horizon, where $\eta=-r$ but first make some assumptions regarding the nature of the source. We assume that the physical size of the system $D(\eta)$ is uniformly bounded by $D_0$ on all cosmological slices $\eta=\text{const}$ and much smaller than the cosmological radius: $D_0\ll1/H$. This means that a binary system will remain compact regardless of the expansion of the universe. Moreover, we assume that the system is stationary at distant past and distant future, namely:
\begin{align}
\mathcal{L}_T T_{\alpha\beta}=0
\end{align}
outside some finite time, where $T$ is the Killing vector field:
\begin{align}
T=-H\Big(\eta\partial_\eta + x\partial_x+ y\partial_y+ z\partial_z\Big).
\end{align}
Ashtekar et al. \cite{ABK2} refer to $T$ as the time translation vector field because its equal to the limit of the time translation Killing vector field for Schwarzschild-de Sitter spacetime for vanishing mass and also its limit for $\Lambda$ going to zero reduces to a time translation in Minkowski spacetime. 

Finally, we use the slow motion approximation, meaning that the velocity of the source is small $v\ll 1$ in $c=1$ units that we work in, and conclude that the solution (\ref{exactsolution}) may be simplified to:
\begin{align}\label{solution}
\chi_{ab}(\eta,\vec x) = \frac{4}{r}\int {d^3 \vec x'} \mathcal{T}_{ab}(\eta_{_\text{ret}},\vec x') + 4\int d^3\vec x' \int_{-\infty}^{\eta_{_\text{ret}}} d\eta' \frac{1}{\eta'} \partial_{\eta'}\mathcal{T}_{ab}(\eta',\vec x'),
\end{align}
where $\eta_{\text{ret}}=\eta-r$. 

\subsection{Expressing retarded solution $\chi_{ab}$ in terms of quadruple moments}
Next goal is to express the retarded solution (\ref{solution}) in terms of the quadruple moments. Again we use the same definition of quadruple moments as in \cite{ABK3}, namely:
\begin{align}
Q_{ab}^{(\rho)}(\eta)&:= \int_\Sigma d^3V \rho(\eta) \bar x_a \bar x_b,\label{massquadruplemoment}\\
Q_{ab}^{(p)}(\eta)&:= \int_\Sigma d^3V \big(p_1(\eta)+p_2(\eta)+p_3(\eta)\big) \bar x_a \bar x_b,\label{pressurequadruplemoment}
\end{align}
where $\Sigma$ is any cosmological surface $\eta=\text{const}$, $d^3V$ denotes proper volume element and $\bar x_a:= -\tfrac{1}{H\eta}x_a$. Matter density $\rho$ and pressure $p_i$ are defined as usual\footnote{Notice that there is no summation over $i$ in the definition of pressure $p_i$.}:
\begin{align}
\rho = T_{\alpha\beta}n^\alpha n^\beta,&& p_i=T^{ab}\partial_a x_i \partial_b x_i.
\end{align}
To rewrite the traced-reversed, rescaled perturbation $\chi_{ab}$ in terms of the quadruple moments one uses the similar strategy as in Minkowski spacetime.  Both terms in (\ref{solution}) consist of the integral of spatial components of the stress-energy tensor $\mathcal{T}_{ab}$ of the source. Using the conservation of the stress-energy tensor, $\bar \nabla^\alpha{T_{\alpha\beta}}=0$ one can express the integral in terms of the second derivative of energy density and use definitions (\ref{massquadruplemoment}) and (\ref{pressurequadruplemoment}) to find\footnote{For a full derivation see \cite{ABK3}.}:
\begin{align}
\chi_{ab}&=-\tfrac{2H}{r}\eta_\text{ret}[\ddot Q_{ab}^{(\rho)}+2H\dot Q_{ab}^{(\rho)}+H\dot Q_{ab}^{(p)}+2H^2Q_{ab}^{(p)}]({\eta_\text{ret}})\nonumber\\
&-2H\int^{\eta_\text{ret}}_{-\infty} \frac{d\eta'}{\eta'}\partial_\eta' \bigg(\eta'[\ddot Q_{ab}^{(\rho)}+2H\dot Q_{ab}^{(\rho)}+H\dot Q_{ab}^{(p)}+2H^2 Q_{ab}^{(p)}](\eta')\bigg)
\end{align}
where the dot corresponds to the lie derivative with respect to the time translation Killing vector field $T$.

In this work our main goal is to find the generalized quadruple formula for the energy of gravitational radiation passing through the horizon. However, we will also examine its limit for the vanishing cosmological constant $\Lambda$ and check weather it recovers the well-known quadruple formula from Minkowski spacetime \cite{Wald}. To do so, we will have to move to a different coordinate system since $(\eta, \vec x)$ coordinates are not suitable to take such limit. The metric (\ref{etametric}) is manifestly not well-defined when $H\rightarrow 0$. Therefore, we will use coordinates $(t, r,\theta, \varphi)$ (or at times $(t,\vec x)$), defined by:
\begin{align}
\eta= -\tfrac{1}{H}e^{-Ht}, && x = r \cos\theta, && y = r \sin\theta\cos\varphi, &&z = r\sin\theta\sin\varphi,
\end{align}
in which the metric tensor $\bar g_{\alpha\beta}$ takes the following form:
\begin{align}
\bar g_{\alpha\beta} dx^{\alpha}dx^\beta = -dt^2 +e^{2Ht}(dr^2+r^2d\theta^2+r^2\sin^2\theta d\varphi^2).
\end{align}
Now, by setting $H=0$ (which is equivalent to $\Lambda=0$) metric coefficients recover those of the Minkowski metric. In this coordinates the time translation Killing vector field $T$ adapted to the rest frame of the source reads:
\begin{align}
T=\partial_t - Hr\partial_r=\partial_t - e^{-Ht}\partial_r,
\end{align}
where the last equality holds on the cosmological horizon $\mathcal{H}$ defined via:
\begin{align}
r=\tfrac{1}{H}e^{-Ht}.
\end{align} 
The trace-reversed, rescaled perturbation transforms to:
\begin{align}\label{rescaledperturbation}
\chi_{ab}&=\tfrac{2}{r}e^{-Ht_{\text{ret}}}[\ddot Q_{ab}^{(\rho)}+2H\dot Q_{ab}^{(\rho)}+H\dot Q_{ab}^{(p)}+2H^2Q_{ab}^{(p)}]({t_\text{ret}})\nonumber\\
&-2H\int^{t_\text{ret}}_{-\infty} dt' [\dddot Q_{ab}^{(\rho)}+3H\ddot Q_{ab}^{(\rho)}+2H^2 \dot Q_{ab}^{(\rho)}+H\ddot Q_{ab}^{(p)}+3H^2\dot Q_{ab}^{(p)}+2H^3Q_{ab}^{(p)}]
\end{align}
where the dot indicates the Lie derivative with respect to $T$ and $t_\text{ret}$ is the retarded time, namely:
\begin{align}
\dot Q_{ab}(t) :&=\mathcal{L}_{T}Q_{ab}(t)= \partial_t Q_{ab}(t) - 2HQ_{ab}(t),\\
t_\text{ret}:&= -\tfrac{1}{H}\ln( e^{-Ht}+ Hr)).
\end{align}
\subsection{$\eta\eta$-component of the trace-reversed, rescaled perturbation}
To express $\chi_{\eta\eta}$ component in terms of the quadruple moments first one needs to rewrite it in terms of the spatial components $\chi_{ab}$ by solving the gauge conditions (\ref{gauge2}):
\begin{align}
\chi_b&= e^{-2Ht}\int^t e^{Ht'}D^a\chi_{ab}dt' + e^{-2Ht}c_b \label{chia}\\
\bar \chi_{\eta\eta} &= e^{-Ht}\int^t ( D^a\chi_a -He^{Ht'}\chi)dt'+e^{-Ht}d.\label{chieta}
\end{align}
It follows from (\ref{exactsolution2}), that $\chi_a$ vanishes on the past boundary $E^+{(i^-)}$ of the future Poincar\'e patch, and therefore $c_b=0$. On the other hand the relation of $\bar \chi_{\eta\eta}$, $\tilde \chi$ and the trace of $\chi_{ab}$ reads:
\begin{align}
\bar \chi_{\eta\eta} = \tilde \chi - \chi.
\end{align}
Since both, $\tilde \chi$ and $\chi$, are zero on $E^+{(i^-)}$, it follows that the integration constant $d$ also vanishes. The vanishing of integration constants $c_a$ and $d$ is consistent with the no-incoming radiation assumption. Using eq. (\ref{chia}) and (\ref{chieta}) while cutting of the vanishing terms yields:
\begin{align}\label{chietaeta}
 \bar \chi_{\eta\eta} &=e^{-Ht}\int^t \bigg(  e^{-2Ht'}\int^{t'}  e^{Ht''}D^aD^b\chi_{ab}dt''  -He^{Ht'}\chi\bigg)dt'.
\end{align}
Before calculating $\chi_{\eta\eta}$ it is convenient to carry out the integral in expression (\ref{rescaledperturbation}) for $\chi_{ab}$ and separate the terms of order $\mathcal{O}(H^3)$ (or equivalently $\mathcal{O}(1/r^3)$) and higher, namely:
\begin{align}\label{chi2ndorder}
\chi_{ab}&=\tfrac{2}{r}e^{-Ht_{\text{ret}}}[\ddot Q_{ab}^{(\rho)}+2H\dot Q_{ab}^{(\rho)}+H\dot Q_{ab}^{(p)}+2H^2Q_{ab}^{(p)}]({t_\text{ret}})\nonumber\\
&-2H\int^{t_\text{ret}}_{-\infty} dt' [\dddot Q_{ab}^{(\rho)}+3H\ddot Q_{ab}^{(\rho)}+2H^2 \dot Q_{ab}^{(\rho)}+H\ddot Q_{ab}^{(p)}+3H^2\dot Q_{ab}^{(p)}+2H^3Q_{ab}^{(p)}]\nonumber\\
&=\tfrac{2}{r}e^{-Ht}[\partial_t^2 Q_{ab}^{(\rho)}-2H\partial_t Q_{ab}^{(\rho)}+H\partial_tQ_{ab}^{(p)}]({t_\text{ret}})+2H^2 [\partial_t Q_{ab}^{(\rho)}]({t_\text{ret}})+\mathcal{O}(H^3).
\end{align}
Consequently, we plug in the above to (\ref{chietaeta}) and after some tedious calculations find the expression for $\chi_{\eta\eta}$ in terms of the quadruple moments:
\begin{align}
\bar  \chi_{\eta\eta} &=-2(H+\tfrac{1}{r})^2[\partial_t Q^{(\rho)}]({t_\text{ret}})+\tfrac{2}{r}({H}+\tfrac{2}{r}){\tilde x^a\tilde x^b}[\partial_t Q_{ab}^{(\rho)}]({t_\text{ret}})\nonumber\\
&-{2H}t\tfrac{\tilde x^a\tilde x^b}{r(e^{-Ht}+Hr)}[\partial_t^2 Q_{ab}^{(\rho)}]({t_\text{ret}})+{2}(H+\tfrac{1}{r})e^{-Ht}\tfrac{\tilde x^a \tilde x^b}{e^{-Ht}+Hr}[\partial_t^2 Q_{ab}^{(\rho)}]({t_\text{ret}})\nonumber\\
&+{2H}(H+\tfrac{1}{r})\tfrac{\tilde x^a \tilde x^b}{(e^{-Ht}+Hr)}[\partial_tQ_{ab}^{(p)}]({t_\text{ret}})+{2}(\tfrac{1}{r^2}-H^2)\tfrac{\tilde x^a \tilde x^b}{e^{-Ht}+Hr}[\partial_tQ_{ab}^{(\rho)}]({t_\text{ret}})+\mathcal{O}(H^3).
\end{align}
where $\tilde x_a = x_a/r$. The last step is to calculate partial derivative of $\chi_{\eta\eta}$ on the horizon $\mathcal{H}$, namely:
\begin{align}\label{Tchietaeta}
T^\mu\partial_\mu\bar \chi_{\eta\eta}&=-{8H^2}[\partial_t^2 Q^{(\rho)}]({t_\text{ret}})+{6H^2}{\tilde x^a\tilde x^b}[\partial_t^2 Q_{ab}^{(\rho)}]({t_\text{ret}})+{2H}{\tilde x^a \tilde x^b}[\partial_t^3 Q_{ab}^{(\rho)}]({t_\text{ret}})\nonumber\\
&+{2H^2}{\tilde x^a \tilde x^b}[\partial_t^2Q_{ab}^{(p)}]({t_\text{ret}}),
\end{align}
where we dropped out the terms of order $\mathcal{O}(H^3)$ and higher since they will not contribute to the leading order corrections to Einstein's quadruple formula\footnote{We have shown a full derivation of expression (\ref{Tchietaeta}) in Appendix \ref{etaalphachi}.}.

\section{Energy flux through a null surface}
The general formula for the energy flux through a null surface derived by V. Chandrasekaran et al. reads \cite{CFP}:
\begin{align}\label{energy1}
E_\ell &= \tfrac{1}{8\pi}\int_{\Delta \mathcal{N}} d^3V \bigg(\sigma_{AB}\sigma^{AB}-\tfrac{1}{2}\theta^2 \bigg)
\end{align}
where $\sigma_{AB}$ and $\theta$ are shear and expansion, respectively, of a time-translation $\ell$ whereas $\Delta \mathcal{N}$ is the null surface with the proper volume form $d^3V$. We will use eq. (\ref{energy1}) to evaluate the expression of the energy flux through the null surface which is the cosmological horizon $\mathcal{H}$ for the vector field $\ell=T$. The second fundamental form of the null surface $\mathcal{H}$ reads:
\begin{align}
K_{ij}&= \tfrac{1}{2}\mathcal{L}_T q_{ij}
\end{align}
and since $\ell^i K_{ij}=0$ it can be written as $K_{AB}$. One can uniquely decompose $K_{AB}$ into traceless shear $\sigma_{AB}$ and expansion $\theta$, namely:
\begin{align}
K_{AB}&= \tfrac{1}{2} \theta q_{AB}+\sigma_{AB}
\end{align}
where $q_{AB}$ is the induced metric on the space-like section $S$ of $\mathcal{H}$. Consequently, we first find shear and expansion of the time translation $T$:
\begin{align}\label{shear}
\sigma_{AB}&= \tfrac{1}{2}T^{\mu}\partial_\mu \tilde \gamma_{AB}-\tfrac{1}{4}q_{AB}q^{CD}T^{\mu}\partial_\mu \tilde \gamma_{CD},\\\label{expansion}
\theta &=\tfrac{1}{2} q^{AB}T^\mu\partial_\mu \tilde \gamma_{AB},
\end{align}
where $\tilde \gamma_{\alpha\beta}$ is a metric perturbation for which cosmological horizon $\mathcal{H}$ defined by $r=\exp({-Ht})/H$ remains null. Therefore, to use eq. (\ref{energy1}) we need to implement the following gauge fixing:
\begin{align}
\tilde g_{\mu\nu}= \bar g_{\alpha\beta}+ \mathcal{L}_\xi \bar g_{\alpha\beta}
\end{align}
satisfying:
\begin{align}\label{gaugeForNull}
T^\mu \bar g_{\mu j}=0,
\end{align}
which ensures that the vector field $T$ and horizon $\mathcal{H}$ are null. The components of the perturbation $\tilde \gamma_{AB}$ take the following form:
\begin{align}\label{g1}
\tilde \gamma_{\theta\theta}&= \gamma_{\theta\theta} +\mathcal{L}_{\xi}\bar g_{\theta\theta},\\\label{g2}
\tilde \gamma_{\theta\varphi}&= \gamma_{\theta\varphi}+\mathcal{L}_{\xi}\bar g_{\theta\varphi},\\\label{g3}
\tilde \gamma_{\varphi\varphi}&= \gamma_{\varphi\varphi}+\mathcal{L}_{\xi}\bar g_{\varphi\varphi}.
\end{align}
%A simple calculation yields:
%\begin{align}
%\sigma_{AB}\sigma^{AB}-\tfrac{1}{2}\theta^2&= \tfrac{1}{4}\bigg[ q^{AC}q^{BD}(T^\mu\partial_\mu\tilde \gamma_{AB})(T^\nu\partial_\nu\tilde\gamma_{CD}) -(q^{AB}T^\mu\partial_\mu\tilde\gamma_{AB})^2 \bigg]. 
%\end{align}
Furthermore, we make use of the expression for shear (\ref{shear}) and expansion (\ref{expansion}) to write the energy flux formula (\ref{energy1}) in terms of perturbation $\tilde \gamma_{AB}$:
\begin{align}\label{energy2b}
E_T&= \tfrac{1}{16\pi}\int_{\mathcal{H}} d^3V  q^{\theta\theta}q^{\varphi\varphi}\bigg( (T^\mu\partial_\mu\tilde \gamma_{\theta \varphi})^2-T^\nu\partial_\nu \tilde \gamma_{\theta\theta}T^\mu\partial_\mu \tilde \gamma_{\varphi\varphi}\bigg).
\end{align}
%Notice that eq. (\ref{energy2b}) holds only for a surface $\Delta N$ which is null. Therefore, to ensure that the cosmological horizon $\mathcal{H}$ remains null in the perturbed de Sitter spacetime we implement the following gauge fixing:
%\begin{align}
%\tilde g_{\mu\nu}= \bar g_{\alpha\beta}+ \mathcal{L}_\xi \bar g_{\alpha\beta}
%\end{align}
%satisfying:
%\begin{align}
%T^\mu \bar g_{\mu j}=0.
%\end{align}
%The components of the perturbation $\tilde \gamma_{AB}$ take the following form:
%\begin{align}
%\tilde \gamma_{\theta\theta}&= \gamma_{\theta\theta} +\mathcal{L}_{\xi}\bar g_{\theta\theta},\\
%\tilde \gamma_{\theta\varphi}&= \gamma_{\theta\varphi}+\mathcal{L}_{\xi}\bar g_{\theta\varphi},\\
%\tilde \gamma_{\varphi\varphi}&= \gamma_{\varphi\varphi}+\mathcal{L}_{\xi}\bar g_{\varphi\varphi}.
%\end{align}
Next, using eq. (\ref{g1}), (\ref{g2}) and (\ref{g3}) and then splitting the energy flux into three parts, namely $E_T^1$ (which comes from the gauge fixing), $E_T^0$ and the higher order terms yields:
\begin{align}\label{energy2}
E_T&= \tfrac{1}{16\pi}\int_{\mathcal{H}} d^3V  q^{\theta\theta}q^{\varphi\varphi}\bigg( (T^\mu\partial_\mu(\gamma_{\theta\varphi}+\mathcal{L}_{\xi}\bar g_{\theta\varphi}))^2-T^\nu\partial_\nu (\gamma_{\theta\theta} +\mathcal{L}_{\xi}\bar g_{\theta\theta})T^\mu\partial_\mu(\gamma_{\varphi\varphi}+\mathcal{L}_{\xi}\bar g_{\varphi\varphi})\bigg)\nonumber\\
&=\tfrac{1}{16\pi}\int_{\mathcal{H}} d^3V  q^{\theta\theta}q^{\varphi\varphi}\bigg( 2T^\mu\partial_\mu \gamma_{\theta \varphi}T^\mu\partial_\mu \mathcal{L}_{\xi}g_{\theta\varphi}-T^\nu\partial_\nu \gamma_{\theta\theta}T^\mu\partial_\mu \mathcal{L}_{\xi}g_{\varphi\varphi}-T^\mu\partial_\mu \mathcal{L}_{\xi}g_{\theta\theta}T^\mu\partial_\mu\gamma_{\varphi\varphi}\bigg)\nonumber\\
&+ \tfrac{1}{16\pi}\int_{\mathcal{H}} d^3V  q^{\theta\theta}q^{\varphi\varphi}\bigg( (T^\mu\partial_\mu \gamma_{\theta \varphi})^2-T^\nu\partial_\nu \gamma_{\theta\theta}T^\mu\partial_\mu \gamma_{\varphi\varphi}\bigg)+\mathcal{O}(H^2)\nonumber\\
&=: E_T^1+E_T^0+\mathcal{O}(H^2)
\end{align}
where the terms with Lie derivatives may be written explicitly as\footnote{For a more detailed calculations see Appendix \ref{Gauge fixing}.}:
\begin{align}
T^\mu\partial_\mu\mathcal{L}_{\xi}\bar g_{\theta\theta}&=\tfrac{8}{H}\sin^2\theta\chi_{xx} +\tfrac{8}{H}\cos^2\theta\sin^2\varphi\chi_{zz}+\tfrac{8}{H}\cos^2\theta\cos^2\varphi\chi_{yy}\nonumber\\
&-\tfrac{16}{H}\cos\theta\sin\theta\cos\varphi\chi_{xy}-\tfrac{16}{H}\cos\theta\sin\theta\sin\varphi\chi_{xz}+\tfrac{16}{H}\cos^2\theta\sin\varphi\cos\varphi\chi_{yz}\nonumber\\
T^\mu\partial_\mu \mathcal{L}_{\xi}\bar g_{\theta\varphi}&=\tfrac{8}{H}\sin\theta\cos\theta\sin\varphi\cos\varphi\chi_{zz}-8\tfrac{1}{H}\sin\theta\cos\theta\sin\varphi\cos\varphi\chi_{yy}\nonumber\\
&+\tfrac{8}{H}\chi_{xy}\sin^2\theta\sin\varphi-\tfrac{8}{H}\chi_{xz}\sin^2\theta\cos\varphi+\tfrac{8}{H}\chi_{yz}\sin\theta\cos\theta(\cos^2\varphi-\sin^2\varphi))\nonumber\\
T^\mu\partial_\mu\mathcal{L}_{\xi}\bar g_{\varphi\varphi}&=-\tfrac{16}{H}\sin^2\theta\sin\varphi\cos\varphi\chi_{yz}+\tfrac{8}{H}\chi_{zz}\sin^2\theta\cos^2\varphi+\tfrac{8}{H}\chi_{yy}\sin^2\theta\sin^2\varphi\nonumber
\end{align}

%&^&

We want to express energy flux through the cosmological horizon $\mathcal{H}$, parametrized by $(t,\theta,\varphi)$, in terms of the quadruple moments, however first we need to move from the physical perturbation $\gamma_{\alpha\beta}$ to the trace-reversed, rescaled $\chi_{\alpha\beta}$ in the $(\eta,x,y,z)$ chart, namely:
\begin{align}
\gamma_{\theta\theta}&=\tfrac{r^2}{2}e^{2Ht}\bigg[\bar \chi_{\eta\eta}+\chi_{xx}(\sin^2\theta-\cos^2\theta)+ \chi_{yy}(-\sin^2\theta+\cos^2\theta(\cos^2\varphi-\sin^2\varphi))\nonumber\\
&+ \chi_{zz} (-\sin^2\theta-\cos^2\theta(\cos^2\varphi-\sin^2\varphi))-4\bar\chi_{xy} \sin\theta \cos\theta\cos\varphi\nonumber\\
&-4\chi_{xz} \sin\theta \cos\theta\sin\varphi+4\chi_{yz} \cos\theta\sin\varphi \cos\theta\cos\varphi \bigg],\nonumber\\
\gamma_{\varphi\varphi}&=\tfrac{r^2}{2}e^{2Ht}\sin^2\theta\bigg[ \bar \chi_{\eta\eta}- \chi_{xx}+(\bar \chi_{yy}- \chi_{zz})(\sin^2\varphi-\cos^2\varphi)-4\chi_{yz} \sin\varphi\cos\varphi \bigg],\nonumber\\
\gamma_{\theta\varphi}&= {r^2}e^{2Ht}\bigg[(- \chi_{yy}+ \chi_{zz}) \sin\theta\cos\theta\sin\varphi \cos\varphi+\chi_{xy} \sin^2\theta\sin\varphi-\chi_{xz} \sin^2\theta\cos\varphi\nonumber\\
&+\chi_{yz}\sin\theta\cos\theta(-\sin^2\varphi+\cos^2\varphi)\bigg].\nonumber
\end{align}
Plugging the above into the energy flux formula (\ref{energy2}) and integrating over the angles (where possible) yields:
\begin{align}\label{energy3}
E^0_T&=\tfrac{1}{48H^2}\int dt   \bigg((T^\mu\partial_\mu \chi_{xx})^2+(T^\mu\partial_\mu \chi_{yy})^2+(T^\mu\partial_\mu \chi_{zz})^2+4(T^\mu\partial_\mu \chi_{xy})^2+4(T^\mu\partial_\mu \chi_{yz})^2\nonumber\\
&+4(T^\mu\partial_\mu \chi_{xz})^2-2T^\mu\partial_\mu \chi_{xx}T^\nu\partial_\nu \chi_{yy}-2T^\mu\partial_\mu \chi_{yy}T^\nu\partial_\nu \chi_{zz}-2T^\mu\partial_\mu \chi_{xx}T^\nu\partial_\nu \chi_{zz}\bigg)\nonumber\\
&-\tfrac{1}{64\pi H^2}\int dt  \int_0^\pi d\theta\int_0^{2\pi}d\varphi  \sin\theta T^\nu\partial_\nu\bar \chi_{\eta\eta}\bigg( T^\mu\partial_\mu\bar \chi_{\eta\eta}-\tfrac{2}{r^2} x^ax^bT^\mu\partial_\mu\chi_{ab}\bigg),\\
E^1_T&=\tfrac{4}{15H}\int dt \bigg(2\chi_{xx}T^\mu\partial_\mu\chi_{xx}+2 \chi_{yy}T^\mu\partial_\mu\chi_{yy}+2\chi_{zz}T^\mu\partial_\mu \chi_{zz}+6\chi_{xy}T^\mu\partial_\mu\chi_{xy}  \nonumber\\
&+6 \chi_{xz}T^\mu\partial_\mu\chi_{xz} +6\chi_{yz}T^\mu\partial_\mu\chi_{yz}-\chi_{zz}T^\mu\partial_\mu\chi_{xx}-\chi_{yy}T^\mu\partial_\mu\chi_{xx}\nonumber\\
&-\chi_{xx}T^\mu\partial_\mu\chi_{yy}-\chi_{zz}T^\mu\partial_\mu\chi_{yy}-\chi_{xx}T^\mu\partial_\mu\chi_{zz}-\chi_{yy}T^\mu\partial_\mu\chi_{zz}\bigg).\label{energy4}
\end{align}
Notice that the overall factor in the above expressions is of a different power in $H$. We will be interested in finding leading order corrections in $\sqrt{\Lambda}$ to the Einstein's quadruple formula, therefore in $E_T^0$ under the integral we will only focus on the terms of order $\mathcal{O}(H^3)$ and lower, whereas for $E_T^1$ of order $\mathcal{O}(H^2)$, the higher order terms will be ignored.

\section{A generalization of the Einstein's quadruple formula for the energy of gravitational radiation}
The quadruple nature of gravitational waves is manifest in both terms of the energy flux formula, (\ref{energy3}) and (\ref{energy4}). The spatial components $\chi_{ab}$ of the trace-reversed, rescaled perturbation may be expressed in terms of the quadruple moments $Q_{ab}^{(\rho)}$ and $Q_{ab}^{(p)}$, eq. (\ref{rescaledperturbation}). 

Now we are ready to rewrite the generalized quadruple formula (\ref{energy2}) in terms of the mass and pressure quadruple moments:
\begin{align}\label{energyfinal0}
E_T&=\tfrac{2}{15}\int dt   \Bigg[(\partial_t^3 Q_{xx}^{(\rho)})^2+(\partial_t^3 Q_{yy}^{(\rho)})^2+(\partial_t^3 Q_{zz}^{(\rho)})^2+3(\partial_t^3 Q_{xy}^{(\rho)})^2+3(\partial_t^3 Q_{xz}^{(\rho)})^2+3(\partial_t^3 Q_{yz}^{(\rho)})^2\nonumber\\
&-(\partial_t^3 Q_{xx}^{(\rho)})(\partial_t^3 Q_{yy}^{(\rho)})-(\partial_t^3 Q_{xx}^{(\rho)})(\partial_t^3 Q_{zz}^{(\rho)})-(\partial_t^3 Q_{yy}^{(\rho)})(\partial_t^3 Q_{zz}^{(\rho)})\nonumber\\
&+2H\partial_t^3 Q_{xx}^{(\rho)}\Big(\partial_t^2Q_{xx}^{(p)}+7\partial_t^2 Q_{xx}^{(\rho)}\Big)+2H\partial_t^3 Q_{yy}^{(\rho)}\Big(\partial_t^2Q_{yy}^{(p)}+7\partial_t^2 Q_{yy}^{(\rho)}\Big)\nonumber\\
&+2H\partial_t^3 Q_{zz}^{(\rho)}\Big(\partial_t^2Q_{zz}^{(p)}+7\partial_t^2 Q_{zz}^{(\rho)}\Big)+6H\partial_t^3 Q_{xy}^{(\rho)}\Big(\partial_t^2Q_{xy}^{(p)}+7\partial_t^2 Q_{xy}^{(\rho)}\Big)\nonumber\\
&+6H\partial_t^3 Q_{xz}^{(\rho)}\Big(\partial_t^2Q_{xz}^{(p)}+7\partial_t^2 Q_{xz}^{(\rho)}\Big)+6H\partial_t^3 Q_{yz}^{(\rho)}\Big(\partial_t^2Q_{yz}^{(p)}+7\partial_t^2 Q_{yz}^{(\rho)}\Big)\nonumber\\
&-H\partial_t^3 Q_{xx}^{(\rho)}\Big(\partial_t^2Q_{yy}^{(p)}+7\partial_t^2 Q_{yy}^{(\rho)}\Big)-H\partial_t^3 Q_{yy}^{(\rho)}\Big(\partial_t^2Q_{xx}^{(p)}+7\partial_t^2 Q_{xx}^{(\rho)}\Big)\nonumber\\
&-H\partial_t^3 Q_{xx}^{(\rho)}\Big(\partial_t^2Q_{zz}^{(p)}+7\partial_t^2 Q_{zz}^{(\rho)}\Big)-H\partial_t^3 Q_{zz}^{(\rho)}\Big(\partial_t^2Q_{xx}^{(p)}+7\partial_t^2 Q_{xx}^{(\rho)}\Big)\nonumber\\
&-H\partial_t^3 Q_{yy}^{(\rho)}\Big(\partial_t^2Q_{zz}^{(p)}+7\partial_t^2 Q_{zz}^{(\rho)}\Big)-H\partial_t^3 Q_{zz}^{(\rho)}\Big(\partial_t^2Q_{yy}^{(p)}+7\partial_t^2 Q_{yy}^{(\rho)}\Big)\Bigg]({t_\text{ret}})+\mathcal{O}(H^2).
\end{align}
At first glance the above expression seems quite complicated, however it can be written in a more compact and readable form.  That requires introducing:
\begin{align}
q^{(i)}_{ab}:=Q^{(i)}_{ab}-\tfrac{1}{3}\mathring q_{ab}Q^{(i)},
\end{align}
where $i$ stands for $\rho$ (mass quadruple moment) or $p$ (pressure quadruple moment) and $Q^{(i)}= \mathring q^{ab}Q^{(i)}_{ab}$.  It follows that $q^{(i)}$ is traceless. Consequently, the quadruple formula for the energy of gravitational radiation in de Sitter spacetime simplifies to:
 \begin{align}\label{energyfinal}
E_T&=\tfrac{1}{5}\int dt\sum_{i,j=1}^3\Bigg[\bigg( \tfrac{d^3q^{(\rho)}_{ij}}{dt^3}\bigg)^2+2H\bigg( \tfrac{d^3q^{(\rho)}_{ij}}{dt^3}\Big(\tfrac{d^2q^{(p)}_{ij}}{dt^2}+7\tfrac{d^2q^{(\rho)}_{ij}}{dt^2}\Big)\bigg)\Bigg](t_{\text{ret}})+\mathcal{O}(H^2).
%&=\tfrac{1}{5}\int dt\sum_{i,j=1}^3\Bigg[\bigg( \tfrac{d^3q^{(\rho)}_{ij}}{dt^3}\bigg)\bigg(\tfrac{d^3q^{(\rho)}_{ij}}{dt^3}+2H \Big(\tfrac{d^2q^{(p)}_{ij}}{dt^2}+7\tfrac{d^2q^{(\rho)}_{ij}}{dt^2}\Big)\bigg)\Bigg](t_{\text{ret}})+\mathcal{O}(H^2)
\end{align}
The final expression for the energy carried away by gravitational waves through the horizon consists of not only the third time derivatives of the mass quadruple moment (as in Minkowski case) but also the first order corrections with the second time derivatives of the mass and pressure quadruple moments and is evaluated at retarded time $t_\text{ret}= -\ln( e^{-Ht}+ Hr)/H \text{ } =\text{ } 2t$, where the last equality holds on the horizon $\mathcal{H}$. 

Since the energy flux (\ref{energyfinal}) is written in the $(t,\vec x)$ chart, taking the limit for vanishing cosmological constant is very straightforward as it only requires setting $H=0$ (recall that $H=\sqrt{\Lambda/3}$), namely:
\begin{align}\label{energylimit}
E_T&=\tfrac{1}{5}\int dt \sum_{i,j=1}^3\Bigg[\bigg( \tfrac{d^3q^{(\rho)}_{ij}}{dt^3}\bigg)^2\Bigg](t_{\text{ret}}).
\end{align}
The limit is a well-known Einstein's quadruple formula in Minkowski spacetime\footnote{Note that in \cite{Wald} the quadruple moments are defined as $Q_{ab}=3\int T_{00}x_ax_b$, and therefore the overall factor in (\ref{energylimit}), considering such definition, is not 1/5 but 1/45.} \cite{Wald}. 

\section{Summary}
We have considered the first order time changing matter source producing the gravitational radiation in de Sitter spacetime. The cosmological Killing horizon was chosen as a generalization of the conformal boundary $\mathcal{I}^+$  used in the radiation theory in the Minkowski spacetime.  We used the retarded solutions to the linearized Einstein's equations on the background de Sitter spacetime provided in \cite{ABK3}.  However, in addition to the spatial components of the solution we needed the $\eta\eta$-component of the traced-reversed, rescaled perturbation. 

We considered the general formula for the energy flux passing through a null surface (\ref{energy1}) which was formulated in \cite{CFP}. However, to apply this formula we needed to introduce  a gauge transformation, ensuring that the horizon remains a null surface with respect to the perturbed geometry, and the Killing vector continues to be null  (\ref{gaugeForNull}). This lead us to the calculation of the energy radiated by the compact source (\ref{energy2}) and also (\ref{energy3}) together with (\ref{energy4}) written in terms of the perturbation $\bar \chi_{\alpha\beta}$.  

Finally,  we calculated the generalized quadruple formula in terms of the mass and pressure quadruple moments $Q_{ab}^{(\rho)}$ and $Q_{ab}^{(p)}$, respectively (\ref{energyfinal0}), and wrote it explicitly up to the first order in $\sqrt{\Lambda}$.
The obtained expression is quite lengthy and to put it in a more compact and readable form we have introduced traceless quadruple moments $q_{ab}^{(i)}$. Indeed, we were able to write the formula (\ref{energyfinal0}) in an elegant way (\ref{energyfinal}). The zeroth order term recovers the famous Einstein's quadruple formula (\ref{energylimit}) for the perturbed Minkowski spacetime, whereas the first order term is a new correction.
   
\vspace{8mm}

{\noindent{\bf Acknowledgements:}}
DDR and JL thank Abhay Ashtekar and Maciej Kolanowski for the discussions.
This work was supported by OPUS 2017/27/B/ST2/02806 of the Polish National Science Centre.

\appendix
\section{Expressing $\eta\alpha$-components of the trace-reversed, rescaled perturbation in terms of the quadruple moments}\label{etaalphachi}
Finding the expression for $\chi_{\eta\eta}$ (or its derivative with respect to the vector field $T$) and $\chi_{\eta a}$ in terms of quadruple moments is crucial for finding leading order corrections to Einstein's quadruple formula. Start with rewriting the expression (\ref{chietaeta}) for $\eta\eta$-component that we obtained from gauge conditions (\ref{gauge2}), namely:
\begin{align}\label{chietaetaapp}
 \bar \chi_{\eta\eta} &=e^{-Ht}\int^t \bigg(  e^{-2Ht'}D^aD^b\int^{t'}  e^{Ht''}\chi_{ab}dt''  -He^{Ht'}\chi\bigg)dt'.
\end{align}
For the calculations in this section it is convenient to write partial derivatives with respect to $t$, $r$ and $x^a$ of quadruple moments, which are a function of retarded time $t_\text{ret}$, as partial derivatives w.r.t. $t$ calculated at $t_\text{ret}$:
\begin{align}\label{partialt1}
\partial_t [Q_{ab}^{(i)}({t_\text{ret}})]&=[\partial_t Q_{ab}^{(i)}]({t_\text{ret}})\partial_t({t_\text{ret}})=[\partial_t Q_{ab}^{(i)}]({t_\text{ret}})/(Hre^{Ht}+1),\\
\partial_r [Q_{ab}^{(i)}({t_\text{ret}})]&=[\partial_t Q_{ab}^{(i)}]({t_\text{ret}})\partial_r({t_\text{ret}})=-[\partial_t Q_{ab}^{(i)}]({t_\text{ret}})/(Hr+e^{-Ht}),\\\label{partialDat}
D^a [Q_{ab}^{(i)}({t_\text{ret}})] &= [\partial_t Q_{ab}^{(i)}]({t_\text{ret}})D^a({t_\text{ret}})= -\tfrac{x^a}{r(e^{-Ht}+Hr)}[\partial_t Q_{ab}^{(i)}]({t_\text{ret}}),%=  -\tfrac{x^a}{r}e^{Ht}\partial_t[ Q_{ab}^{(\rho)}]({t_\text{ret}})
\end{align}
where $i$ stands for $\rho$ or $p$ and:
\begin{align}
t_\text{ret}:&= -\tfrac{1}{H}\ln( e^{-Ht}+ Hr)).
\end{align}
To find the energy flux formula (\ref{energy3}) we don't need to find explicit expression for $\chi_{\eta\eta}$, but its derivative with respect to vector field $T$, namely:
\begin{align}\label{Tpartialchietaeta}
T^\mu\partial_\mu \chi_{\eta\eta}&= (\partial_t-e^{-Ht}\partial_r)\Bigg[e^{-Ht}\int^t \bigg(  e^{-2Ht'}\int^{t'}  e^{Ht''}D^aD^b\chi_{ab}dt''  -He^{Ht'}\chi\bigg)dt'\Bigg]\nonumber\\
&= -He^{-Ht}\int^t \bigg(  e^{-2Ht'}\int^{t'}  e^{Ht''}D^aD^b\chi_{ab}dt''  -He^{Ht'}\chi\bigg)dt'\nonumber\\
&+  e^{-3Ht}\int^{t'}  e^{Ht''}D^aD^b\chi_{ab}dt''  -H\chi\nonumber\\
&- e^{-2Ht}\int^t \bigg(  e^{-2Ht'}\int^{t'}  e^{Ht''}\partial_rD^aD^b\chi_{ab}dt''  -He^{Ht'}\chi\bigg)dt'
\end{align}
We will calculate each term of the last equality separately, starting with:
\begin{align}
-He^{-Ht}\int^t \bigg(  e^{-2Ht'}\int^{t'}  e^{Ht''}D^aD^b\chi_{ab}dt''  -He^{Ht'}\chi\bigg)dt'
\end{align}
and neglecting $\mathcal{O}(H^3)$ terms. Notice that the second term in the above has a factor of $H^2$ in front of the trace of $\chi_{ab}$ which itself is of order $H$. Consequently, the lowest order will be the term $\mathcal{O}(H^3)$, which we neglect in our calculations. Therefore we obtain:
\begin{align}
-He^{-Ht}\int^t &\bigg(  e^{-2Ht'}\int^{t'}  e^{Ht''}D^aD^b\chi_{ab}dt''  -He^{Ht'}\chi\bigg)dt'\nonumber\\
&=-He^{-Ht}\int^t  e^{-2Ht'}\int^{t'}  e^{Ht''}D^aD^b\chi_{ab}dt'' dt' + \mathcal{O}(H^3).
\end{align}
Before we proceed with our derivation recall that:
\begin{align}\label{chiabappendix}
\chi_{ab}&=\tfrac{2}{r}e^{-Ht_{\text{ret}}}[\ddot Q_{ab}^{(\rho)}+2H\dot Q_{ab}^{(\rho)}+H\dot Q_{ab}^{(p)}+2H^2Q_{ab}^{(p)}]({t_\text{ret}})\nonumber\\
&-2H\int^{t_\text{ret}}_{-\infty} dt' [\dddot Q_{ab}^{(\rho)}+3H\ddot Q_{ab}^{(\rho)}+2H^2 \dot Q_{ab}^{(\rho)}+H\ddot Q_{ab}^{(p)}+3H^2\dot Q_{ab}^{(p)}+2H^3Q_{ab}^{(p)}]\nonumber\\
&=\tfrac{2}{r}e^{-Ht}[\partial_t^2Q_{ab}^{(\rho)}-2H\partial_t Q_{ab}^{(\rho)}+H\partial_t Q_{ab}^{(p)}]({t_\text{ret}})\nonumber\\
&+2H[H\partial_t Q_{ab}^{(\rho)}-2H^2 Q_{ab}^{(\rho)}+H^2Q_{ab}^{(p)}]({t_\text{ret}}).
\end{align}
Again, the last two terms will be neglected because they consist of the overall factor of $H^3$. We then act with derivative $D^a$ on expression (\ref{chiabappendix}) to obtain:
\begin{align}\label{Dchi}
D^a\chi_{ab}&=-\tfrac{2x^a}{r^3}e^{-Ht}[\partial_t^2Q_{ab}^{(\rho)}]({t_\text{ret}})-2\tfrac{x^a}{r^2(1+Hre^{Ht})}[\partial_t^3Q_{ab}^{(\rho)}-2H\partial_t^2 Q_{ab}^{(\rho)}+H\partial_t^2 Q_{ab}^{(p)}]({t_\text{ret}})\nonumber\\
&-2H^2e^{Ht}\tfrac{x^a}{r(1+Hre^{Ht})}[\partial_t^2 Q_{ab}^{(\rho)}]({t_\text{ret}}) +\mathcal{O}(H^3).
\end{align}
Then, we find the second derivative of $\chi_{ab}$:
\begin{align}
D^bD^a\chi_{ab}&=-2\tfrac{1}{r^2(1+Hre^{Ht})}[\partial_t^3Q^{(\rho)}]({t_\text{ret}})+2\tfrac{x^ax^b(3Hre^{Ht}+2)}{r^4(1+Hre^{Ht})^2}[\partial_t^3Q_{ab}^{(\rho)}]({t_\text{ret}})\nonumber\\
&+2\tfrac{x^ax^b}{r^4(1+Hre^{Ht})}[\partial_t^3Q_{ab}^{(\rho)}]({t_\text{ret}})+2H^2e^{2Ht}\tfrac{x^ax^b}{r^2(1+Hre^{Ht})^2}[\partial_t^3 Q_{ab}^{(\rho)}]({t_\text{ret}})\nonumber\\
&+2\tfrac{x^ax^b}{r^3(1+Hre^{Ht})^2}e^{Ht}[\partial_t^4Q_{ab}^{(\rho)}-2H\partial_t^3 Q_{ab}^{(\rho)}+H\partial_t^3 Q_{ab}^{(p)}]({t_\text{ret}})+\mathcal{O}(H^3).\nonumber\\
\end{align}
In the next step we calculate the integral of the above expression, namely:
\begin{align}
\int^t e^{Ht'}D^bD^a\chi_{ab}&=-2\tfrac{1}{r^2(1+Hre^{Ht})}[\partial_t^3Q^{(\rho)}]({t_\text{ret}})+2\tfrac{x^ax^b(3Hre^{Ht}+2)}{r^4(1+Hre^{Ht})}e^{Ht}[\partial_t^2Q_{ab}^{(\rho)}]({t_\text{ret}})\nonumber\\
&+2\tfrac{x^ax^b}{r^4}e^{Ht}[\partial_t^2Q_{ab}^{(\rho)}]({t_\text{ret}})+2H^2e^{3Ht}\tfrac{x^ax^b}{r^2(1+Hre^{Ht})}[\partial_t^2 Q_{ab}^{(\rho)}]({t_\text{ret}})\nonumber\\
&+2\tfrac{x^ax^b}{r^3(1+Hre^{Ht})}e^{2Ht}[\partial_t^3Q_{ab}^{(\rho)}-2H\partial_t^2 Q_{ab}^{(\rho)}+H\partial_{t'}^2 Q_{ab}^{(p)}]({t_\text{ret}})\nonumber\\
&+2\int^t\tfrac{1}{r^2(1+Hre^{Ht'})}[\partial_{t'}^3Q^{(\rho)}]({t'_\text{ret}})dt'-2\int^t\partial_{t'}\Big(\tfrac{x^ax^b}{r^4}e^{Ht'}\Big)[\partial_{t'}^2Q_{ab}^{(\rho)}]({t'_\text{ret}})dt'\nonumber\\
&-\int^t\partial_{t'}\Big(2\tfrac{x^ax^b(3Hre^{Ht'}+2)}{r^4(1+Hre^{Ht'})}e^{Ht'}\Big)[\partial_{t'}^2Q_{ab}^{(\rho)}]({t'_\text{ret}})dt'\nonumber\\
&-2\int^t\partial_{t'}\Big(\tfrac{x^ax^b}{r^3(1+Hre^{Ht'})}e^{2Ht'}\Big)[\partial_t'^3Q_{ab}^{(\rho)}-2H\partial_{t'}^2 Q_{ab}^{(\rho)}+H\partial_{t'}^2 Q_{ab}^{(p)}]({t'_\text{ret}})dt'\nonumber\\
&-2H^2\int^t\partial_{t'}\Big(e^{3Ht'}\tfrac{x^ax^b}{r^2(1+Hre^{Ht'})}\Big)[\partial_{t'}^2 Q_{ab}^{(\rho)}]({t'_\text{ret}})dt'+\mathcal{O}(H^3),
\end{align}
where we used integration by parts. Notice that derivatives with respect to $t'$ under the integrands will produce terms proportional to $H$. In other words, they increase the order of $H$. Having in mind that $r=e^{-Ht}/H$ on horizon one sees that all but one of those integrands will consist the terms of order lower than $H^3$, that is:
\begin{align}
-2\int^t&\partial_{t'}\Big(\tfrac{x^ax^b}{r^3(1+Hre^{Ht'})}e^{2Ht'}\Big)[\partial_t'^3Q_{ab}^{(\rho)}]({t'_\text{ret}})dt'\nonumber\\
&=-2H\int^t\tfrac{x^ax^b(Hre^{Ht'}+2)}{r^3(1+Hre^{Ht'})^2}e^{2Ht'}[\partial_t'^3Q_{ab}^{(\rho)}]({t'_\text{ret}})dt'\nonumber\\
&=-2H\int^t\tfrac{x^ax^b(Hre^{Ht'}+2)}{r^3(1+Hre^{Ht'})}e^{2Ht'}\partial_{t'}[\partial_t'^2Q_{ab}^{(\rho)}]({t'_\text{ret}})dt'\nonumber\\
&=-2H\tfrac{x^ax^b(Hre^{Ht}+2)}{r^3(1+Hre^{Ht})}e^{2Ht}[\partial_t^2Q_{ab}^{(\rho)}]({t_\text{ret}})dt+\mathcal{O}(H^3),
\end{align}
where in the second equality we used identity (\ref{partialt1}) and in the last one we again integrated by parts neglecting higher order terms. Consequently, we have:
\begin{align}\label{int1}
\int^t e^{Ht}D^bD^a\chi_{ab}&=-2\tfrac{1}{r^2}[\partial_t^2Q^{(\rho)}]({t_\text{ret}})+2\tfrac{x^ax^b(3Hre^{Ht}+2)}{r^4(1+Hre^{Ht})}e^{Ht}[\partial_t^2Q_{ab}^{(\rho)}]({t_\text{ret}})\nonumber\\
&+2\tfrac{x^ax^b}{r^4}e^{Ht}[\partial_t^2Q_{ab}^{(\rho)}]({t_\text{ret}})+2H^2e^{3Ht}\tfrac{x^ax^b}{r^2(1+Hre^{Ht})}[\partial_t^2 Q_{ab}^{(\rho)}]({t_\text{ret}})\nonumber\\
&+2\tfrac{x^ax^b}{r^3(1+Hre^{Ht})}e^{2Ht}[\partial_t^3Q_{ab}^{(\rho)}-2H\partial_t^2 Q_{ab}^{(\rho)}+H\partial_{t}^2 Q_{ab}^{(p)}]({t_\text{ret}})\nonumber\\
&-2H\tfrac{x^ax^b(Hre^{Ht}+2)}{r^3(1+Hre^{Ht})}e^{2Ht}[\partial_t^2Q_{ab}^{(\rho)}]({t_\text{ret}})+\mathcal{O}(H^3).
\end{align}
In the final step of this derivation we calculate the second integrant of the above and find its value on the horizon:
\begin{align}
-He^{-Ht}\int^t &\Big(  e^{-2Ht'}\int^{t'}  e^{Ht''}D^aD^b\chi_{ab}dt''  -He^{Ht'}\chi\Big)dt'=-2H^2{\tilde x^a\tilde x^b}[\partial_t^2Q_{ab}^{(\rho)}]({t_\text{ret}})+\mathcal{O}(H^3).
\end{align}
Next we calculate the second line in the last equality of (\ref{Tpartialchietaeta}) taking the trace of (\ref{chiabappendix}) and using expression (\ref{int1}):
\begin{align}
 e^{-3Ht}\int^{t'}  e^{Ht''}D^aD^b\chi_{ab}dt''  -H\chi&=-2\tfrac{e^{-3Ht}}{r^2}[\partial_t^2Q^{(\rho)}]({t_\text{ret}})+2\tfrac{x^ax^b(3Hre^{Ht}+2)}{r^4(1+Hre^{Ht})}e^{-2Ht}[\partial_t^2Q_{ab}^{(\rho)}]({t_\text{ret}})\nonumber\\
&+2\tfrac{x^ax^b}{r^4}e^{-2Ht}[\partial_t^2Q_{ab}^{(\rho)}]({t_\text{ret}})+2H^2\tfrac{x^ax^b}{r^2(1+Hre^{Ht})}[\partial_t^2 Q_{ab}^{(\rho)}]({t_\text{ret}})\nonumber\\
&+2\tfrac{x^ax^b}{r^3(1+Hre^{Ht})}e^{-Ht}[\partial_t^3Q_{ab}^{(\rho)}-2H\partial_t^2 Q_{ab}^{(\rho)}+H\partial_{t}^2 Q_{ab}^{(p)}]({t_\text{ret}})\nonumber\\
&-2H\tfrac{x^ax^b(Hre^{Ht}+2)}{r^3(1+Hre^{Ht})}e^{-Ht}[\partial_t^2Q_{ab}^{(\rho)}]({t_\text{ret}})-\tfrac{2H}{r}e^{-Ht}[\partial_t^2Q^{(\rho)}]({t_\text{ret}})\nonumber\\
&+\mathcal{O}(H^3).\nonumber\\
\end{align}
Finally, we calculate the last term of (\ref{Tpartialchietaeta}):
\begin{align}
- e^{-2Ht}\int^t& \bigg(  e^{-2Ht'}\int^{t'}  e^{Ht''}\partial_rD^aD^b\chi_{ab}dt''  -He^{Ht'}\partial_r\chi\bigg)dt'\nonumber\\
&=-2\tfrac{e^{-2Ht}}{r^2}[\partial_t^2Q^{(\rho)}]({t_\text{ret}})+2\tfrac{\tilde x^a\tilde x^b(3Hre^{Ht}+2)}{r^2(1+Hre^{Ht})}e^{-2Ht}[\partial_t^2Q_{ab}^{(\rho)}]({t_\text{ret}})\nonumber\\
&+2\tfrac{\tilde x^a\tilde x^b}{r^2}e^{-2Ht}[\partial_t^2Q_{ab}^{(\rho)}]({t_\text{ret}})+2\tfrac{\tilde x^a\tilde x^b}{r(1+Hre^{Ht})}e^{-Ht}[\partial_t^3Q_{ab}^{(\rho)}-2H\partial_t^2 Q_{ab}^{(\rho)}+H\partial_{t}^2 Q_{ab}^{(p)}]({t_\text{ret}})\nonumber\\
&+2H^2e^{2Ht}\tfrac{\tilde x^a\tilde x^b}{(1+Hre^{Ht})}e^{-2Ht}[\partial_t^2 Q_{ab}^{(\rho)}]({t_\text{ret}})+2\tfrac{\tilde x^a\tilde x^b(3Hre^{Ht}+1)}{r^2(1+Hre^{Ht})}e^{-2Ht}[\partial_t^2Q_{ab}^{(\rho)}]({t_\text{ret}})\nonumber\\
&-2H\tfrac{\tilde x^a\tilde x^b(Hre^{Ht}+3)}{r(1+Hre^{Ht})}e^{-Ht}[\partial_t^2Q_{ab}^{(\rho)}]({t_\text{ret}})-2H\tfrac{\tilde x^a\tilde x^b}{r(1+Hre^{Ht})}e^{-Ht}[\partial_t^2Q_{ab}^{(\rho)}]({t_\text{ret}})\nonumber\\
&-\tfrac{2H}{r}e^{-Ht}[\partial_t^2Q^{(\rho)}]({t_\text{ret}})+\mathcal{O}(H^3).
\end{align}
Taking it all together and setting $r=e^{-Ht}/H$ as we are interested in the result on the horizon we find:
\begin{align}
T^\mu\partial_\mu \chi_{\eta\eta}&=-8H^2[\partial_t^2Q^{(\rho)}]({t_\text{ret}})+6H^2{\tilde x^a\tilde x^b}[\partial_t^2Q_{ab}^{(\rho)}]({t_\text{ret}})\nonumber\\
&+2H{\tilde x^a\tilde x^b}[\partial_t^3Q_{ab}^{(\rho)}+H\partial_{t}^2 Q_{ab}^{(p)}]({t_\text{ret}}).
\end{align}
Gauge condition (\ref{gauge2}) lets us also express the $\eta a$-component of the retarded solution in terms of $\chi_{ab}$ and consequently quadruple moments:
\begin{align}
\bar \chi_{\eta b}=\chi_b&= e^{-2Ht}\int^t e^{Ht'}D^a\chi_{ab}dt'\nonumber\\
&=-{2}H^2 t\tfrac{x^a}{r}\partial_t[\partial_t Q_{ab}^{(\rho)}]({t_\text{ret}})-{2}(H+\tfrac{1}{r})(1-Ht)\tfrac{x^a}{r}\partial_t[\partial_t Q_{ab}^{(\rho)}]({t_\text{ret}})\nonumber\\
&-{2H}(H+\tfrac{1}{r})\tfrac{x^a}{r}\partial_t[-2 Q_{ab}^{(\rho)}+Q_{ab}^{(p)}]({t_\text{ret}})-{2}(H+\tfrac{1}{r})\tfrac{x^a}{r^2}\partial_t[Q_{ab}^{(\rho)}]({t_\text{ret}})\nonumber\\
&=-{2}H^2 t\tfrac{x^a}{r(Hre^{Ht}+1)}[\partial_t^2 Q_{ab}^{(\rho)}]({t_\text{ret}})-{2}(H+\tfrac{1}{r})(1-Ht)\tfrac{x^a}{r(Hre^{Ht}+1)}[\partial_t^2 Q_{ab}^{(\rho)}]({t_\text{ret}})\nonumber\\
&-{2H}(H+\tfrac{1}{r})\tfrac{x^a}{r(Hre^{Ht}+1)}[-2 \partial_t Q_{ab}^{(\rho)}+\partial_t Q_{ab}^{(p)}]({t_\text{ret}})-{2}(H+\tfrac{1}{r})\tfrac{x^a}{r^2(Hre^{Ht}+1)}[\partial_t Q_{ab}^{(\rho)}]({t_\text{ret}})\nonumber\\
&=-2H\tilde x^a[\partial_t^2 Q_{ab}^{(\rho)}]({t_\text{ret}})+\mathcal{O}(H^2)\end{align}
where we used (\ref{Dchi}) and focused on terms linear in $H$ since higher order terms will not contribute to the first order corrections to quadruple formula. Notice that last equality holds on the horizon.

\section{Gauge fixing}\label{Gauge fixing}
Energy flux formula (\ref{energy1}) holds for any null surface $\Delta \mathcal{N}$. Since we want to find the flux going through the cosmological horizon $\mathcal{H}$ in a perturbed de Sitter spacetime we need to make sure that the perturbation does not effect the null character of the horizon. To satisfy that condition we will introduce the following gauge fixing:
\begin{align}\label{nullgauge}
T^\mu \tilde g_{\mu a} = 0
\end{align}
where $T$ is tangent and null on the horizon $\mathcal{H}$ and metric tensor $ \tilde g_{\mu \nu}$ is of a form:
\begin{align}
 \tilde g_{\mu \nu} = g_{\mu\nu} +  \mathcal{L}_\xi g_{\mu\nu}.
\end{align}
It is convenient to introduce coordinates adapted to the horizon:
\begin{align}
v&=\eta+r, \nonumber\\
u&=\eta-r. \nonumber
\end{align}
The transformed de Sitter metric takes the form:
\begin{align}
\bar g_{\mu\nu}dx^\mu dx^\nu &= \tfrac{1}{H^2(u+v)^2}\bigg(-4dudv+(v-u)^2(d\theta^2+\sin^2\theta d\varphi^2)\bigg),
\end{align}
whereas the surface $v=0$ is the cosmological horizon $\mathcal H$. In $(u,v,\theta,\varphi)$ chart the gauge condition (\ref{nullgauge}) can be written as three equations satisfied on the horizon:
\begin{align}
 \gamma_{uu} +\mathcal{L}_{\xi}\bar g_{uu}&= 0,\\
 \gamma_{u\theta} +\mathcal{L}_{\xi}\bar g_{u\theta}&= 0,\\
 \gamma_{u\varphi}+\mathcal{L}_{\xi}\bar g_{u\varphi} &= 0.
\end{align}
Writing out explicitly the Lie derivatives yields:
\begin{align}
\gamma_{uu}  - \tfrac{4}{H^2u^2}\partial_u \xi^v&= 0\\
 \gamma_{u\theta} +\tfrac{1}{H^2u^2}\big( u^2\partial_u \xi^\theta-2\partial_\theta \xi^v\big)&= 0\\
 \gamma_{u\varphi}+\tfrac{1}{H^2u^2}\big( u^2 \sin^2\theta \partial_u \xi^\varphi-2\partial_\varphi \xi^v \big)&= 0
\end{align}
Solving for $\xi$ gives:
\begin{align}
  \xi^v&=\tfrac{1}{4}H^2\int^uu'^2 \gamma_{uu}du'\nonumber\\
   \xi^\theta &= \int^u\Big(2\tfrac{1}{u'^2}\partial_\theta \xi^v-{H^2} \gamma_{u\theta}\Big)du'\nonumber\\
   \xi^\varphi&=\tfrac{1}{\sin^2\theta} \int^u\Big(2\tfrac{1}{u'^2}\partial_\varphi \xi^v-{H^2} \gamma_{u\varphi}\Big)du'\nonumber
\end{align}
Next, we transform vector $\xi$ to the $(t, r, \theta, \varphi)$ chart:
\begin{align}
\xi &=\tfrac{1}{2}e^{Ht} \xi^v \partial_t +\tfrac{1}{2}\xi^v \partial_r+ \xi^\theta\partial_\theta + \xi^\varphi \partial_\varphi\nonumber\\
\end{align}
as well as the perturbation $\gamma_{ua}$:
\begin{align}
\gamma_{uu}&= \sin^2\theta\cos^2\varphi\chi_{yy}+\cos^2\theta\chi_{xx}+\sin^2\theta\sin^2\varphi\chi_{zz}+2\sin\theta\cos\theta\cos\varphi\chi_{xy}\nonumber\\
&+2\sin\theta\cos\theta\sin\varphi\chi_{xz}+2\sin^2\theta\sin\varphi\cos\varphi\chi_{yz}\nonumber\\
\gamma_{u\theta}&={\tfrac{1}{H}\sin\theta\cos\theta\chi_{xx} -\tfrac{1}{H}\sin\theta\cos\theta\sin^2\varphi\chi_{zz}-\tfrac{1}{H}\sin\theta\cos\theta\cos^2\varphi\chi_{yy}}\nonumber\\
&{-\tfrac{1}{H}(\cos^2\theta-\sin^2\theta)\cos\varphi\chi_{xy}-\tfrac{1}{H}(\cos^2\theta-\sin^2\theta)\sin\varphi\chi_{xz}-\tfrac{2}{H}\sin\theta\cos\theta\sin\varphi\cos\varphi\chi_{yz}}\nonumber\\
\gamma_{u\varphi}&={-\tfrac{1}{H}\chi_{zz}\sin^2\theta\sin\varphi\cos\varphi+\tfrac{1}{H}\chi_{yy}\sin^2\theta\sin\varphi\cos\varphi}\nonumber\\
&{+\tfrac{1}{H}\chi_{xy}\sin\theta\cos\theta\sin\varphi-\tfrac{1}{H}\chi^{(1)}_{xz}\sin\theta\cos\theta\cos\varphi-\tfrac{1}{H}\chi_{yz}\sin^2\theta(\cos^2\varphi-\sin^2\varphi)}\nonumber\\
\end{align}
where we used the leading order corespondance between $\chi_{\eta \alpha}$ and $\chi_{ab}$ (derived in Appendix \ref{etaalphachi}). For the energy expression (\ref{energy1}) we will need the angular components of the perturbation, namely:
\begin{align}
\tilde \gamma_{\theta\theta}&= \gamma_{\theta\theta} +\mathcal{L}_{\xi}\bar g_{\theta\theta},\\
\tilde \gamma_{\theta\varphi}&= \gamma_{\theta\varphi}+\mathcal{L}_{\xi}\bar g_{\theta\varphi},\\
\tilde \gamma_{\varphi\varphi}&= \gamma_{\varphi\varphi}+\mathcal{L}_{\xi}\bar g_{\varphi\varphi}.
\end{align}
which written in terms of the vector field $\xi$ read:
\begin{align}
\tilde \gamma_{\theta\theta}&= \gamma_{\theta\theta} +2He^{2Ht}r^2\xi^t+2e^{2Ht}r\xi^r +2e^{2Ht}r^2\partial_\theta \xi^\theta\nonumber\\
\tilde \gamma_{\theta\varphi}&= \gamma_{\theta\varphi}+e^{2Ht}r^2\partial_\varphi \xi^\theta+e^{2Ht}r^2\sin^2\theta\partial_\theta \xi^\varphi\nonumber\\
\tilde \gamma_{\varphi\varphi}&= \gamma_{\varphi\varphi}+2He^{2Ht}r^2\sin^2\theta\xi^t+2e^{2Ht}r\sin^2\theta\xi^r+2e^{2Ht}r^2\sin\theta\cos\theta\xi^\theta +2e^{2Ht}r^2\sin^2\theta\partial_\varphi \xi^\varphi\nonumber\\
\end{align}
To calculate $E_T^1$ term of the energy flux (\ref{energy2}) we need to express $T^\mu\partial_\mu\mathcal{L}_{\xi}\bar g_{AB}$ in terms of trace-reversed, rescaled perturbation $\chi_{ab}$, namely:
\begin{align}
T^\mu\partial_\mu\mathcal{L}_{\xi} g_{\theta\theta}&=4H^2e^{2Ht}r^2\xi^t+2He^{2Ht}r^2\partial_t\xi^t+4He^{2Ht}r\xi^r+2e^{2Ht}r\partial_t\xi^r +4He^{2Ht}r^2\partial_\theta \xi^\theta+2e^{2Ht}r^2\partial_t\partial_\theta \xi^\theta\nonumber\\
&-4H^2e^{2Ht}r^2\xi^t-2He^{2Ht}r^2\partial_r\xi^t-2e^{2Ht}\xi^r-2e^{Ht}Hr\partial_r\xi^r -4Hre^{2Ht}r\partial_\theta \xi^\theta-2He^{2Ht}r^3\partial_r\partial_\theta \xi^\theta\nonumber\\
&=\tfrac{8}{H}\bar \gamma_{uu}+2\int^{u(t)}\bar \gamma_{uu}dt'+{8} \int^{u(t)}\partial^2_\theta\bar \gamma_{uu}dt'-8\partial_\theta\bar \gamma_{u\theta}\nonumber\\
&=\tfrac{8}{H}\bar \gamma_{uu}-8\partial_\theta\bar \gamma_{u\theta}\nonumber\\
&=\tfrac{8}{H}\sin^2\theta\chi_{xx} +\tfrac{8}{H}\cos^2\theta\sin^2\varphi\chi_{zz}+\tfrac{8}{H}\cos^2\theta\cos^2\varphi\chi_{yy}\nonumber\\
&-\tfrac{16}{H}\cos\theta\sin\theta\cos\varphi\chi_{xy}-\tfrac{16}{H}\cos\theta\sin\theta\sin\varphi\chi_{xz}+\tfrac{16}{H}\cos^2\theta\sin\varphi\cos\varphi\chi_{yz}\nonumber\\
\end{align}
where we used the fact that $\xi$  is $r$-independent and considered only the terms in the leading order of $H$. Similar calculation yields the other two terms:
\begin{align}
T^\mu\partial_\mu \mathcal{L}_{\xi}\bar g_{\theta\varphi}&=\tfrac{8}{H}\sin\theta\cos\theta\sin\varphi\cos\varphi\chi_{zz}-8\tfrac{1}{H}\sin\theta\cos\theta\sin\varphi\cos\varphi\chi_{yy}\nonumber\\
&+\tfrac{8}{H}\chi_{xy}\sin^2\theta\sin\varphi-\tfrac{8}{H}\chi_{xz}\sin^2\theta\cos\varphi+\tfrac{8}{H}\chi_{yz}\sin\theta\cos\theta(\cos^2\varphi-\sin^2\varphi),\nonumber\\
T^\mu\partial_\mu\mathcal{L}_{\xi}\bar g_{\varphi\varphi}&=-\tfrac{16}{H}\sin^2\theta\sin\varphi\cos\varphi\chi_{yz}+\tfrac{8}{H}\chi_{zz}\sin^2\theta\cos^2\varphi+\tfrac{8}{H}\chi_{yy}\sin^2\theta\sin^2\varphi.\nonumber\\
\end{align}

\newpage
%\clearpage

 \thispagestyle{empty}
\end{document}